\documentclass[showpacs,amsmath,amssymb,twocolumn,superscriptaddress,notitlepage,preprintnumbers,pra]{revtex4-1}
\usepackage{color}
\usepackage{physics}
\usepackage{graphicx}
\usepackage{dcolumn}%

\usepackage{hyperref}
\usepackage{float}
\usepackage{subfigure}
\hypersetup{colorlinks=true, linkcolor=blue, citecolor=blue, urlcolor=black}

\usepackage{amsmath, amsthm, amssymb, bm}

\newtheorem{theorem}{Theorem}

\newcommand{\Ebb}{\mathbb{E}}

\newcommand{\pbra}[1]{\left ( #1 \right)}

\newcommand{\sbra}[1]{\left[ #1 \right]}

\definecolor{wbj}{rgb}{0.4,0.3,0.9}

\begin{document}


\title{Tensor-Network-Assisted Variational Quantum Algorithm}

\author{Junxiang Huang}
\thanks{These two authors contributed equally.}
\affiliation{School of Computer Science, Peking University, Beijing 100871, China}
\affiliation{Center on Frontiers of Computing Studies, Peking University, Beijing 100871, China}
\affiliation{School of Physics, Peking University, Beijing 100871, China}

\author{Wenhao He}
\thanks{These two authors contributed equally.}
\affiliation{Center on Frontiers of Computing Studies, Peking University, Beijing 100871, China}
\affiliation{School of Physics, Peking University, Beijing 100871, China}


\author{Yukun Zhang}
\affiliation{Center on Frontiers of Computing Studies, Peking University, Beijing 100871, China}

\author{Yusen Wu}
\affiliation{Department of Physics, University of Western Australia, Perth, Western Australia 6009, Australia}

\author{Bujiao Wu}
 \email{bujiaowu@gmail.com}
\affiliation{Center on Frontiers of Computing Studies, Peking University, Beijing 100871, China}
\affiliation{Dahlem Center for Complex Quantum Systems, Freie Universit\"{a}t, Berlin 14195, Germany}
 
\author{Xiao Yuan}
\email{xiaoyuan@pku.edu.cn}
\affiliation{Center on Frontiers of Computing Studies, Peking University, Beijing 100871, China}





\date{\today}

\begin{abstract}
Near-term quantum devices generally suffer from shallow circuit depth and hence limited expressivity due
to noise and decoherence. To address this, we propose tensor-network-assisted parametrized quantum circuits,
which concatenate a classical tensor-network operator with a quantum circuit to effectively increase the circuit’s
expressivity without requiring a physically deeper circuit. We present a framework for tensor-network-assisted
variational quantum algorithms that can solve quantum many-body problems using shallower quantum circuits.
We demonstrate the efficiency of this approach by considering two examples of unitary matrix-product operators
and unitary tree tensor networks, showing that they can both be implemented efficiently. Through numerical
simulations, we show that the expressivity of these circuits is greatly enhanced with the assistance of tensor
networks.We apply our method to two-dimensional Ising models and one-dimensional time-crystal Hamiltonian
models with up to 16 qubits and demonstrate that our approach consistently outperforms conventional methods
using shallow quantum circuits.
\end{abstract}

\maketitle


\section{Introduction}


Tensor networks (TNs) and parametrized quantum circuits (PQCs) are powerful tools for representing quantum many-body states in classical and quantum approaches, respectively.
The density-matrix renormalization-group (DMRG) algorithm, based on TNs, has achieved great success in determining ground state properties for one-dimensional systems~\cite{white1992density,white1993density,rommer1997class}.
However, the expressivity of TNs is limited by the area law with limited bond dimensions. Parametrized quantum circuits, on the other hand, offer a more natural representation of quantum states on quantum computers, and many quantum algorithms~\cite{cao2019quantum,mcardle2020quantum} have been proposed to take advantage of this. 
Nevertheless, near-term quantum computers are inherently noisy, which could also limit the circuit depth and expressivity of PQCs. 
Therefore, finding systems with non-trivial entanglement structures, such as strongly correlated matters and molecules, using either TNs or PQCs remains a challenging task.


Tensor networks and PQCs are commonly considered as distinct classical and quantum computation methods, each with its own set of advantages and limitations. 
While TNs are relatively easy to implement, they have limited expressivity due to the area law, while PQCs offer much larger expressivity but are limited by noise and shallow circuit depth. 
Nevertheless, TNs and PQCs have been shown to have close interactions with each other. Parametrized quantum circuits, for example, can be designed as classically unrealizable TNs with exponentially large bond dimensions~\cite{huggins2019towards,foss2021holographic,bauer2020quantum,lubasch2020variational,barratt2021parallel}. 
At the same time, TNs that are classically realizable can represent special unitary operations and be used as a particular type of PQC. This raises the question of whether we can integrate these two methods under a unified framework.



Here we present a framework for tensor-network-assisted variational quantum algorithms. 
Our proposal involves a tensor-network-parametrized quantum circuit (TN-PQC) framework, which consist of a standard PQC with an appended TN unitary operator. 
By augmenting the PQC with the TN unitary operator, which mainly performs classical rotations of the Hamiltonian, the TN-PQC framework can significantly enhance circuit depth and thereby improve expressiveness without requiring the physical implementation of deeper circuits.
We then proceed to examine three key questions pertaining to our framework:
(i) how to design the TN-PQC structure, (ii) optimization strategies for the TN-PQC structure, and (iii) the comparative benefits of this hybrid architecture.
To address (i), we present two examples of unitary matrix product operator (uMPO) and unitary tree tensor network (uTTN) and demonstrate their efficacy.
We address (ii) with various optimization strategies and tackle (iii) through numerical experiments.
We implement our method to numerically estimate the ground energy of the 2D Ising model with 16 qubits and the 1D time crystal Hamiltonian with 11-16 qubits.
We compare the performance of the TN-PQC (uMPO), TN-PQC (uTTN), and VQE algorithms. Our numerical results highlight the significant advantages of TN-PQC methods over conventional methods, with the TN-PQC (uMPO) method exhibiting the best performance, suggesting the benefits of uMPO integration.


This paper is organized as follows. In Sec.~II we provide an overview of TN and VQE. In Sec.~III we present our TN-PQC framework, accompanied by a theoretical analysis of its expressivity, and introduce two optimization strategies. In Sec.~IV we present numerical experiments involving two specific TNs, which confirm the advantages of our proposed framework in terms of expressiveness, reduction in the number of VQE layers, and robustness to noise. We summaize in Sec.~V.

\textit{Related work.}
Several hybrid networks have been proposed to enhance the capabilities of the VQE~\cite{shang2021schr,zhang2022variational}.
Shang \textit{et al}.~\cite{shang2021schr} proposed a hybrid algorithm that embedded the Clifford circuit into the VQE algorithm.
Since Clifford gates are unitary and can be simulated in polynomial time, they can be used to provide a unitary transformation to the Hamiltonian $H$, denoted by $H \rightarrow U_c(\theta) H U_c^\dagger(\theta)$, where $U_c(\theta)$ is selected from the Clifford group, thus leading to a more powerful hybrid \textit{Ansatz}.
Zhang \textit{et al}.~\cite{zhang2022variational} enhanced the capability of the \textit{Ansatz} by combining a shallow parametrized quantum circuit with classical neural networks. Rudolph \textit{et al}.~\cite{rudolph2023synergy} introduce classical tensor networks to assist PQC initialization. Yuan \textit{et al}.~\cite{yuan2021quantum} proposed hybrid tensor networks to enhance quantum simulation.

\section{background}
This section provides an introduction to the concepts that are relevant to this paper, including the variational quantum eigensolver (VQE), tensor networks, unitary tensor networks, and the representation of Hamiltonians.

\subsection{Variational quantum eigensolver (VQE)}
In order to harness the potential of noisy intermediate-scale quantum (NISQ) devices, the variational quantum eigensolver (VQE) has received widespread attention \cite{wang2019accelerated,kandala2017hardware,parrish2019quantum,cerezo2021variational}. 
The VQE relies on the Rayleigh-Ritz variational principle \cite{Rayleigh} to enable quantum computers to optimize the lowest possible expectation value of the trial wavefunction as an approximation to the ground state energy of a quantum system.

As a hybrid quantum-classical approach, VQE algorithms follow the typical paradigm of updating a parametrized quantum circuit using classical optimization methods to minimize a given loss function. 
In recent years, there has been significant development of VQE techniques, both theoretically and experimentally, especially for applications in quantum chemistry and physics within the NISQ era \cite{cerezo2021variational}.

The scalability of NISQ quantum computing is currently hindered by the presence of various types of noise, including coherent and incoherent noise as well as measurement errors, which accumulate as the quantum circuit depth increases. 
As a result, the limited depth of quantum circuits restricts their expressive power, preventing them from capturing nontrivial ground-state entanglement structures for complex quantum systems such as strongly correlated materials or molecules. 
This dilemma can be framed as a trade-off between circuit fidelity and expressivity, where the latter refers to a circuit's ability to generate a sufficient number of quantum states to encompass the solution to a given problem.
To overcome these limitations, classical resources are becoming increasingly necessary, and one such example is the use of tensor networks in hybrid \textit{Ansätze}~\cite{yuan2021quantum} to aid in addressing these recent obstacles to quantum computing.


\subsection{Tensor network}

\begin{figure}[htb]
\centering
  \includegraphics[width=0.5\textwidth]{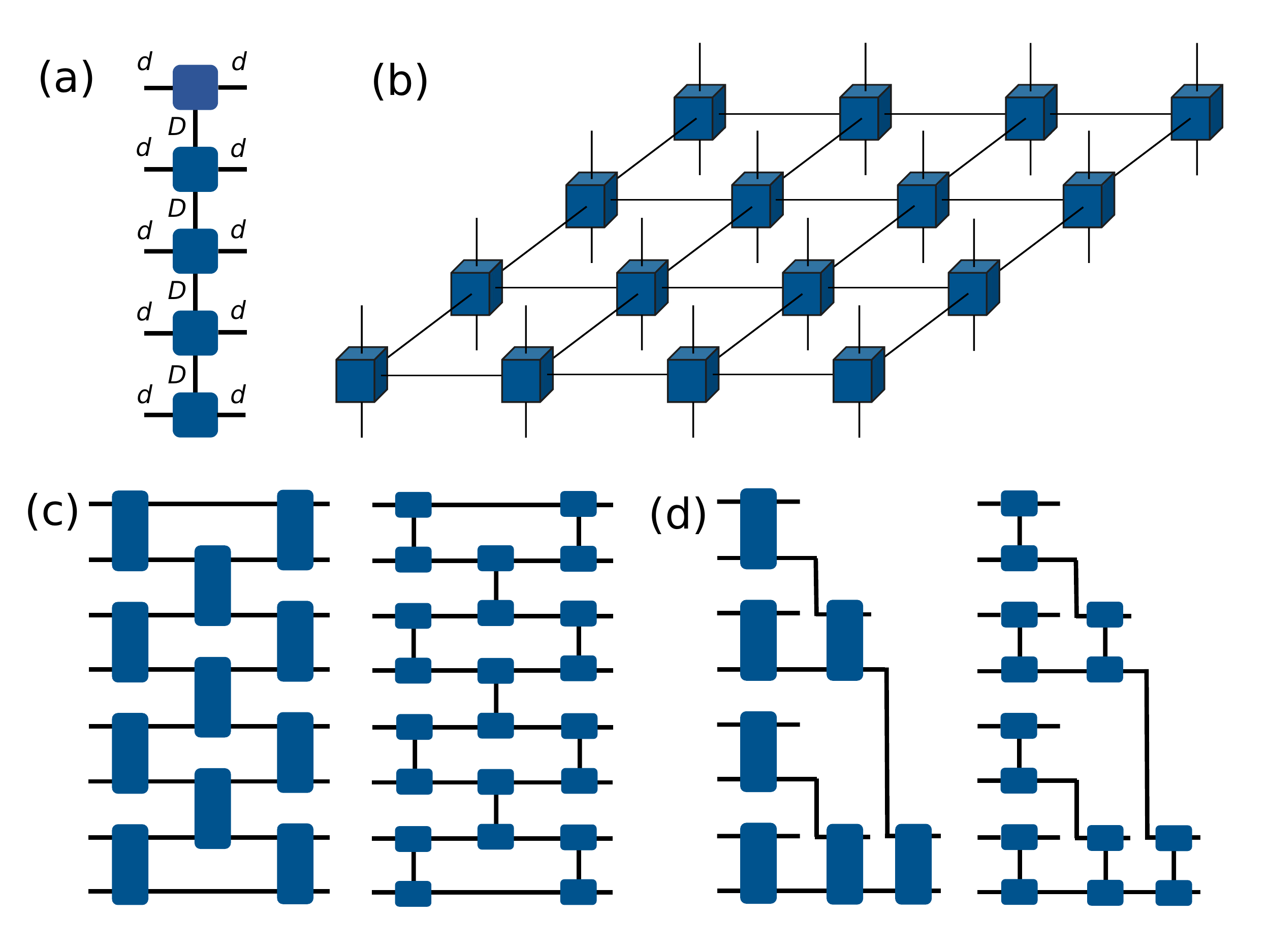}
  \caption{Illustration of several common tensor networks. (a) Matrix-product operator (MPO) with $N=5$. (b) Projected entangled-pair operator (PEPO). (c) Unitary matrix product operator (uMPO) and its equivalent MPO form. The left figure shows the structure of a uMPO, while the right figure decomposes each block on the left using singular value decomposition (SVD), resulting in an MPO with three layers. (d) Structure of a unitary tree tensor network (uTTN) and its SVD form.}
  \label{fig:TN}
\end{figure}

Tensor networks (TNs) are structured tensors constructed from networks of smaller tensors. They are widely used to represent quantum states or operators in quantum many-body systems~\cite{orus2019tensor,shi2006classical}. 
Legs in a network without connection are called dangling legs. 
When all dangling legs represent distinct qubit sites, the TN is called a tensor-network state. 
Alternatively, if all dangling legs are divided into row and column indices, the TN can be defined as a tensor-network operator. In this paper, we focus on the latter case.

There are several types of TN structures, but we will mainly focus on matrix product operators (MPOs), which are used to represent lattice model Hamiltonians, and unitary tensor networks (uTNs).



A matrix product operator (MPO) is a chain of interconnected tensors, each with two dangling legs, as illustrated in Fig.~\ref{fig:TN}(a). 
Each block represents a ($d\times d\times D\times D$) dimensional tensor with order 4, where $d$ is the Hilbert space dimension of each site and $D$ is the bond dimension. 
In this paper, we consider qubit systems, where $d=2$. 
Additionally, the bond dimension $D$ measures the complexity of the contraction of the MPO, which can be estimated with $O(ND^3)$ time complexity, where $N$ is the length of the chain~\cite{ran2020tensor}. 
Figure~\ref{fig:TN}(b) shows a graphical representation of the projected entangled-pair operator (PEPO)\cite{kshetrimayum2019tensor}, where each layer is a 1D MPO with the same form as shown in Fig.~\ref{fig:TN}(a).


Tensor networks can also be used to represent transformations. A unitary tensor network (uTN) is a TN that becomes an identity tensor when contracted with its complex conjugate TN~\cite{haghshenas2021optimization,pollmann2016efficient}.
With the uTN structure, we can express the similarity transformation of a unitary operator $U$ on a Hamiltonian $H$, i.e., $H' = U^\dagger H U$, in tensor network diagrams.
One way to construct a uTN is to use unitary tensors as building blocks, as a network consisting of unitary tensors would automatically be a uTN. 
Following this approach, we can construct several types of uTN structures. 
For example, two well-known TN structures, the MPO and tree tensor network (TTN), can be extended to uTNs.


One uTN structure is derived from an MPO, as shown in Fig.~\ref{fig:TN}(c). 
If a singular value decomposition (or other decomposition methods) is performed on each block, it becomes evident that this uTN originates from an MPO. 
However, it does not possess the global entanglement of a traditional MPO, but only an $O(l)$ local entanglement, where $l$ is the number of layers. 
Therefore, it is better to regard it only as a brick-wall TN with a regular layer structure.

Another uTN structure is shown in Fig.~\ref{fig:TN}(d), which we refer to as a unitary tree tensor network (uTTN). 
It performs coarse-grained processing on the system. 
Unlike the traditional tree structure tensor network where each tensor has three legs, each tensor in the uTTN has four legs, with one leg serving as the output. 
The output of this uTTN consists of different layers of coarse-grained structures.




In addition to quantum states and operations, $k$-local Hamiltonians can also be represented by MPOs~\cite{PhysRevB.95.125125}. 
For instance, the MPO for the transverse Ising model, $H = -J \sum_i (Z_iZ_{i+1} + gX_i)$, can be constructed as $J W^{[1]}W^{[2]}\cdots W^{[n]}$, where $n$ is the number of qubits in the chain and $J$ and $g$ are the parameters in the transverse Ising model. The MPO representation for the transverse Ising model consists of $n$ blocks, each of which is a $3 \times 3$ matrix with $2 \times 2$ matrices as elements, and is denoted by $W^{[i]}$ for $i = 1, 2, \ldots, n$. Specifically,
\begin{eqnarray}
W^{[1]}=\left (\begin{array}{ccc}
I_1 & -Z_1 & g X_1
\end{array}\right), \\
W^{[n]}=\left (\begin{array}{ccc}
g X_n & -Z_n & I_n
\end{array}\right)^{T}, \\
W^{[i]}=\left (\begin{array}{ccc}
I_i & -Z_i & g X_i \\
0 & 0 & Z_i \\
0 & 0 & I_i
\end{array}\right) \\ \text{ for } i = 2, 3, \ldots, n-1,
\end{eqnarray}
where $I_i$, $X_i$, $Y_i$, and $Z_i$ are the identity and Pauli-$X$, Pauli-$Y$, and Pauli-$Z$ operators on the $i$th qubit, respectively. Each $W^{[i]}$ represents a $3 \times 3 \times 2 \times 2$ tensor ($3 \times 2 \times 2$ for $W^{[1]}$ and $W^{[n]}$) with two dangling legs and two contracted legs that connect to the neighboring blocks. The MPO representation of the Hamiltonian serves as a building block for the density-matrix renormalization-group algorithm and provides an intuitive way to use tensor networks to assist the VQE.



\section{The TN-PQC Framework}
\begin{figure*}[htb]
\centering
  \includegraphics[width=\textwidth]{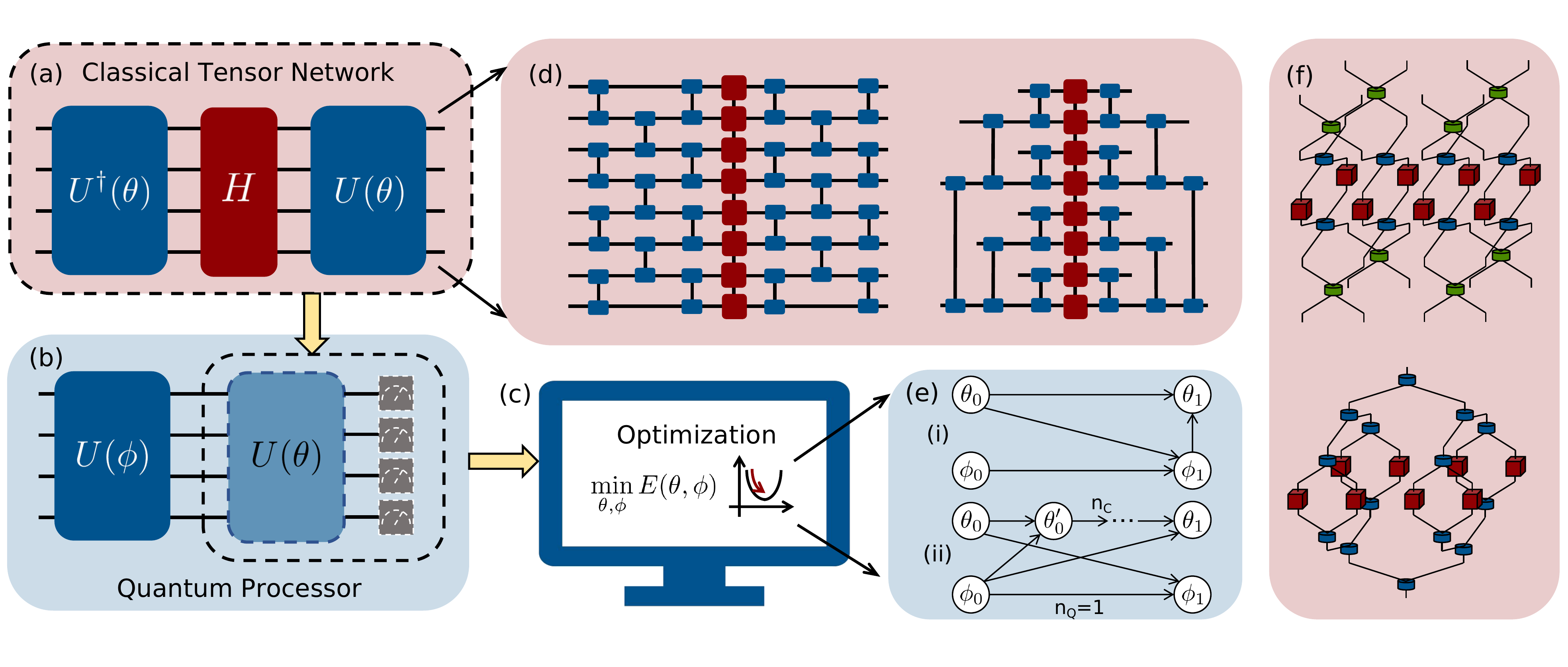}
  \caption{The TN-PQC framework. (a) A classical tensor network can represent part of the parametrized circuits and the Hamiltonian $U^\dagger(\theta) HU(\theta)$. (b) The quantum processor measures the remaining parametrized circuits using an effective Hamiltonian represented by a TN. (c) The gradient descent method is employed to find classical and quantum parameters corresponding to the lowest measurement results. (d) Contraction strategies for the 1D uMPO and uTTN when combined with the Hamiltonian in a TN form. (e) Two optimization strategies for different parameters. (i) The iteration of the quantum parameter $\boldsymbol{\phi}_1$ is obtained by computing the gradient from the previous classical and quantum parameters $\boldsymbol{\theta}_0$ and $\boldsymbol{\phi}_0$, while the parameter $\boldsymbol{\theta}_1$ is obtained from $\boldsymbol{\theta}_0$ and $\boldsymbol{\phi}_1$. (ii) The parameter updating process is similar, but with the introduction of $n_C>1$ intermediate steps between the initial and final classical parameters $\boldsymbol{\theta}_0$ and $\boldsymbol{\theta}_1$ to update classical parameters. (f) The 2D uMPO and uTTN combined with the Hamiltonian in a 2D TN form.}
  \label{fig:framework}
\end{figure*}

The problem of finding the ground state and ground energy of the Hamiltonian $H$ can be expressed as the minimization problem:
\begin{eqnarray}
\arg\min_{\psi} \bra{\psi} H \ket{\psi},
\end{eqnarray}
where different parametrizations of $\psi$ result in different methods.
One example is to set $\ket{\psi} = U (\boldsymbol{\phi})\ket{\bm 0}$, where $U (\boldsymbol{\phi})$ is a PQC with tunable parameters $\boldsymbol{\phi}$. This method is known as a traditional VQE.
In the framework of this article, we set $\ket{\psi} = U (\boldsymbol{\theta})U (\boldsymbol{\phi})\ket{\bm 0}$, which transforms the minimization problem into:
\begin{eqnarray}
\min_{\boldsymbol{\theta}, \boldsymbol{\phi}} \bra{\bm 0}U^\dagger (\boldsymbol{\phi})U^\dagger (\boldsymbol{\theta}) H U (\boldsymbol{\theta})U (\boldsymbol{\phi})\ket{\bm 0},
\label{eq:min_problem}
\end{eqnarray}
where $U (\boldsymbol{\phi})$ represents a quantum circuit with parameters $\boldsymbol{\phi}$, and $U (\boldsymbol{\theta})$ represents a unitary tensor network with parameters $\boldsymbol{\theta}$. By optimizing both classical and quantum parameters, the limited circuit depth of $U (\boldsymbol{\phi})$ can be compensated by the higher expressive power of $U (\boldsymbol{\theta})$. We aim to disentangle Hamiltonian $H$ using the similarity transformation $U (\boldsymbol{\theta})$ and make the VQE circuit more efficient.

When both $U (\boldsymbol{\theta})$ and $H$ are expressed in TNs, the parametrized Hamiltonian $H(\bm \theta) := U^\dagger (\boldsymbol{\theta}) H U (\boldsymbol{\theta})$ in Eq.~\eqref{eq:min_problem} can be considered as a TN with some legs contracted. The entire parametrized TN $H(\bm \theta)$ is referred to as the TN part in the following sections, while $U (\boldsymbol{\phi})\ket{\psi_0}$ in Eq.~\eqref{eq:min_problem} is referred to as the VQE part. When the parameter $\boldsymbol{\theta}$ is fixed, Eq.~\eqref{eq:min_problem} reduces to a traditional VQE method, whereas when $\boldsymbol{\phi}$ is fixed, it becomes a classical parameter optimization method.
The entire process described above is illustrated in Figs.~\ref{fig:framework}(a)-\ref{fig:framework}(c).



One issue to consider is how to choose the PQC $U (\boldsymbol{\phi})$ and the TN $U (\boldsymbol{\theta})$. 
In principle, the structure of the PQC $U (\boldsymbol{\phi})$ can be arbitrary, and in this paper we utilize some existing VQE \textit{Ansätze}~\cite{cerezo2021variational}. However, designing the TN requires more expertise to avoid the explosion of the Pauli decomposition of the TN part $H(\bm\theta)$, since the VQE requires measuring the Pauli basis. 
With the Pauli expansion form of the TN part, the resulting energy can be expressed as a linear combination of expectation values
\begin{eqnarray}
E (\boldsymbol{\theta}, \boldsymbol{\phi}) = \sum_P c_P (\boldsymbol{\theta}) \langle P \rangle_{\boldsymbol{\phi}},
\label{eq:energy}
\end{eqnarray}
where $c_P (\boldsymbol{\theta}) = \text{Tr} [PU^\dagger (\boldsymbol{\theta}) H U (\boldsymbol{\theta})]$ is the Pauli decomposition coefficient of the classical part and the summation iterates over all operators $P$ with non-zero coefficients. The $\langle P \rangle_{\boldsymbol{\phi}}$ denotes the measurement results of the operator $P$ on the PQC, with the subscript indicating the parameter in the PQC. 

Next we will discuss how to choose the TN, the optimization strategy, and the expressivity of this framework.

\subsection{Choosing the TN}

We propose three conditions for selecting an appropriate TN $U (\boldsymbol{\theta})$ in the classical part: 
\begin{enumerate}
\item[(i)] $U^\dagger (\boldsymbol{\theta}) H U (\boldsymbol{\theta})$ must share the same ground state and energy as $H$.
\item[(ii)] Either the number of Pauli decomposition terms remains small, which is a polynomial in terms of the qubit number, or the Pauli operators can be effectively sampled.
\item[(iii)] The coefficients $c_P (\boldsymbol{\theta})$ must be able to be computed efficiently.
\end{enumerate}

Condition (i) can be satisfied by using a unitary $U (\boldsymbol{\theta})$, i.e., a uTN, which is utilized in this paper. 
An alternative approach to fulfill (i) is to use a nonunitary Hermitian $U (\boldsymbol{\theta})$ and dividing by a normalization factor $ \bra{\psi_0}U^\dagger (\boldsymbol{\phi})U^\dagger (\boldsymbol{\theta}) U (\boldsymbol{\theta})U (\boldsymbol{\phi})\ket{\psi_0}$ in the minimization problem \eqref{eq:min_problem} \cite{sokolov2022orders, zhang2022variational}.

Condition (ii) is critical to ensure that the number of Pauli measurements is manageable. Two methods to fulfill condition (ii) are (a) restricting the depth of the uTN $U (\boldsymbol{\theta})$ and (b) designing a specific uTN $U (\boldsymbol{\theta})$ that maps Pauli operators to another single or a few Pauli operators. Both approaches will be discussed below.

For condition (iii), an efficient contraction strategy for $\text{Tr} (U^\dagger (\boldsymbol{\theta}) H U (\boldsymbol{\theta}) P)$, where $P$ is a Pauli operator in the context of a tensor network, is necessary. We will discuss these conditions in specific cases, particularly (ii) and (iii).

    

In Fig.~\ref{fig:framework}(d) we illustrate two structures in the 1D Hamiltonian case, which are derived from the uMPO and uTTN, respectively.

Let us first consider condition (ii) in the uMPO-guided structure. We define an $m$-neighboring Pauli string as a Pauli string with the distance of each of its two nonidentity terms less than $m$. 
For example, an XIXIII string is a 3-neighboring Pauli string, and simple Hamiltonians like the transverse Ising model consist of all 1- or 2-neighboring strings. 
Because each layer of the uMPO only contains blocks that entangle neighboring sites, any 2-neighboring Pauli string will become the superposition of several ($2+2$)-neighboring Pauli strings after commuting with this layer. 
After the similarity transformation, an $l$-layer uTN is transformed into ($2+2l$)-neighboring Pauli strings. 
It is easy to check that the number of general ($2+2l$)-neighboring Pauli strings is limited to $4^{2+2l}$. 
Taking the transverse Ising model as an example, $U^\dagger HU$ only contains less than $n4^{2+2l}$ Pauli strings, where $n$ is the number of qubits. 
When $l\sim\mathcal O(\log n)$, this satisfies condition (ii).

The contraction strategy is as follows: Tensors are divided into groups by row, tensors in each group are contracted first, and then groups are contracted with each other. 
According to the fact that the contraction complexity is equal to $O(D^3)$, where $D$ is the bond dimension, the time complexity for contracting the uTN in the uMPO case is $O(n\times2l\times2^3 + n\times (4^{2l} \times d)^3) = O(n4^{6l})$, where $n$ is the number of qubits, $l$ is the number of layers, and $d$ is the bond dimension for the Hamiltonian. This satisfies condition (iii) as well.

We can apply the results obtained earlier to verify conditions (ii) and (iii) in the uTTN case. 
Since the number of layers equals $l = \log_2 n$ in the uTTN, the number of Pauli strings equals $n4^{2+2l}=O(n^5)$. 
Similarly, the contraction complexity is $O(n^{13})$. 
Both results are polynomial in the number of qubits $n$, indicating that conditions (ii) and (iii) are satisfied.

For a 2D Hamiltonian, we can devise a corresponding 2D uMPO and the coarse-grained structure in the 2D case, as shown in Fig.~\ref{fig:framework}(f). 
The 2D uMPO building block is the same as that for the 1D uMPO, a four-leg unitary tensor. The lower-layer entangles sites aligned in one direction, and the upper layer entangles sites in another. 
We call this structure MPO for the same reason as mentioned in the 1D MPO, and it can be considered a 2D MPO when each block is singular value decomposed.
As in the previous analysis, each site of the 2D Ising model has ($4+2l$)-neighboring Pauli strings. For the 2D uMPO, the number of Pauli strings is limited to a constant $O(4^{4+2l})$ and the time complexity of contraction is $O(n4^{6l})$.
Hence, the 2D uTTN corresponds to a number $O(n^5)$ of Pauli strings and the contraction complexity is also $O(n^{13})$.
Therefore, the 2D cases of these two TNs similarly satisfy conditions (ii) and (iii).


\subsection{Optimization strategy}

Here we propose two optimization strategies for solving the minimization problem in Eq.~\eqref{eq:min_problem}. 
The first strategy is to alternate the optimization of $\boldsymbol{\theta}$ and $\boldsymbol{\phi}$ while keeping the other parameter fixed. 
Figure~\ref{fig:framework}(e i) illustrates each step of this optimization strategy, where the optimization of each parameter is obtained by computing the gradient based on a set of quantum and classical parameter combinations, represented by the arrows in the figure. 

When the classical parameter $\boldsymbol{\theta}$ is frozen, the system reduces to a traditional VQE, and we can perform gradient descent using the parameter shift rule. 
When the quantum parameter $\boldsymbol{\phi}$ is frozen, we can also perform gradient descent using the gradient computed by the formula:
\begin{equation}
\frac{\partial E (\boldsymbol{\theta}, \boldsymbol{\phi})} { \partial \boldsymbol{\theta}} = \sum_P \frac{\partial c_P (\boldsymbol{\theta})}{\partial \boldsymbol{\theta}} \langle P \rangle_{\boldsymbol{\phi}},
\end{equation}
where the coefficient gradient can be computed by the Pauli expansion coefficient of the gradient of the classical part:
\begin{eqnarray}
\frac{\partial c_P (\boldsymbol{\theta})}{\partial \boldsymbol{\theta}} = \text{Tr} (P \frac{\partial U^\dagger (\boldsymbol{\theta}) H U (\boldsymbol{\theta})}{\partial \boldsymbol{\theta}} ).
\end{eqnarray}

The second optimization strategy takes advantage of the fact that $\boldsymbol{\phi}$ and $\boldsymbol{\theta}$ are separated in Eq.\eqref{eq:energy}, allowing these parameters to be updated in parallel. 
Specifically, we start each optimization cycle with initial quantum parameters $\boldsymbol{\phi}_0$ and update it for $n_Q$ steps while keeping initial classical parameters $\boldsymbol{\theta}_0$ constant. 
At the same time, we update $\boldsymbol{\theta}$ for $n_C$ steps while keeping $\boldsymbol{\phi}_0$ constant. 
After both threads have finished, we evaluate Eq.~\eqref{eq:energy} to determine whether the TN-PQC algorithm has converged. 
This evaluation produces a set of measurement results $\langle P \rangle_{\boldsymbol{\phi}}$, which can be reused to compute gradients for the next cycle, as shown in Fig.~\ref{fig:framework}(e ii).

A natural choice for $n_Q$ and $n_C$ is to keep both threads running at the same time, which means setting $n_C/n_Q=\tau_Q/\tau_C$. 
Since NISQ devices are typically slower than classical computers, we can let $n_Q=1$ and choose a relatively large value for $n_C$, which allows classical computers to intervene and help speed up the computation.


\subsection{Expressivity}
In this section, we analyze the expressivity of the TN-PQC framework and demonstrate the trade-off between computational complexity and expressivity. 
Expressivity refers to the ability of a model to represent a wide range of functions. 
Here, we compare the expressivity of the TN-PQC framework with that of a Haar random circuit, which is known to have an intense expressivity. A Haar random circuit can represent any quantum state.

To measure the distance between the states generated by the TN-PQC framework and the Haar random circuit, we use the logarithmic difference of entanglement entropy~\cite{shang2021schr,Iaconis21quantum,liu2018entanglement}. The measure is defined as

\begin{equation}
\Delta_t(\Upsilon) = \log\left(\frac{\mathbb{E}_{\rho\sim\text{Haar}} [{\rm Tr} (\rho^t_{n/2})]}{\mathbb{E}_{\rho\sim \Upsilon }[{\rm Tr} (\rho_{n/2}^t)]}\right),
\label{eq:expressivity}
\end{equation}

where $t\in\mathbb{N}^{+}$ and $\rho_{n/2}$ represents the partial trace of $\rho$ on the first $n/2$ qubits.
From Iaconis \cite{Iaconis21quantum} we have $\mathbb{E}_{\rho\sim\text{Haar}} [{\rm Tr} (\rho^t_{n/2})] = \mathbb{E}_{\rho\sim \mu_k }[{\rm Tr} (\rho_{n/2}^t)]$, where $\mu_k$ is the $k$-design. 
Reference~\cite{liu2018entanglement} gives the bounds for $\Delta_t(\Upsilon)$, where $\Upsilon$ is an $\varepsilon$ approximation of the $t$-design, as shown in Appendix~\ref{app:expressivity}.
Hence we expect that the distribution of the ansatz forms an $\varepsilon$-approximation $t$-design and $\Delta_t$ is negatively correlated to $\varepsilon$.
 For the TN-PQC method, the distribution $\Upsilon$ is chosen as the distribution of the \textit{ansatz} space.
 We will utilize the approximations of $\Delta_2$ and $\Delta_3$ to compare numerically how well the TN-PQC and VQE methods approximate the 2-design and 3-design, respectively, in the next section.



\section{Numerical results and Comparison}

In this section we present a series of numerical experiments to demonstrate the effectiveness and versatility of our TN-PQC method. 
First, we investigate the expressiveness of the TN-PQC and VQE methods by measuring their logarithmic difference. 
Second, we analyze the scalability and accuracy advantages of the TN-PQC method over classical methods. 
Finally, we evaluate the noise resistance of the TN-PQC method by assessing its performance for noise models. 
Through these experiments, we aim to provide a comprehensive evaluation of the TN-PQC method's capabilities and potential applications in practical quantum computing tasks.

\subsection{Outperformance of the TN-PQC method in expressivity}

\begin{figure*}[htb]
\centering
\includegraphics[width=1.0\textwidth]{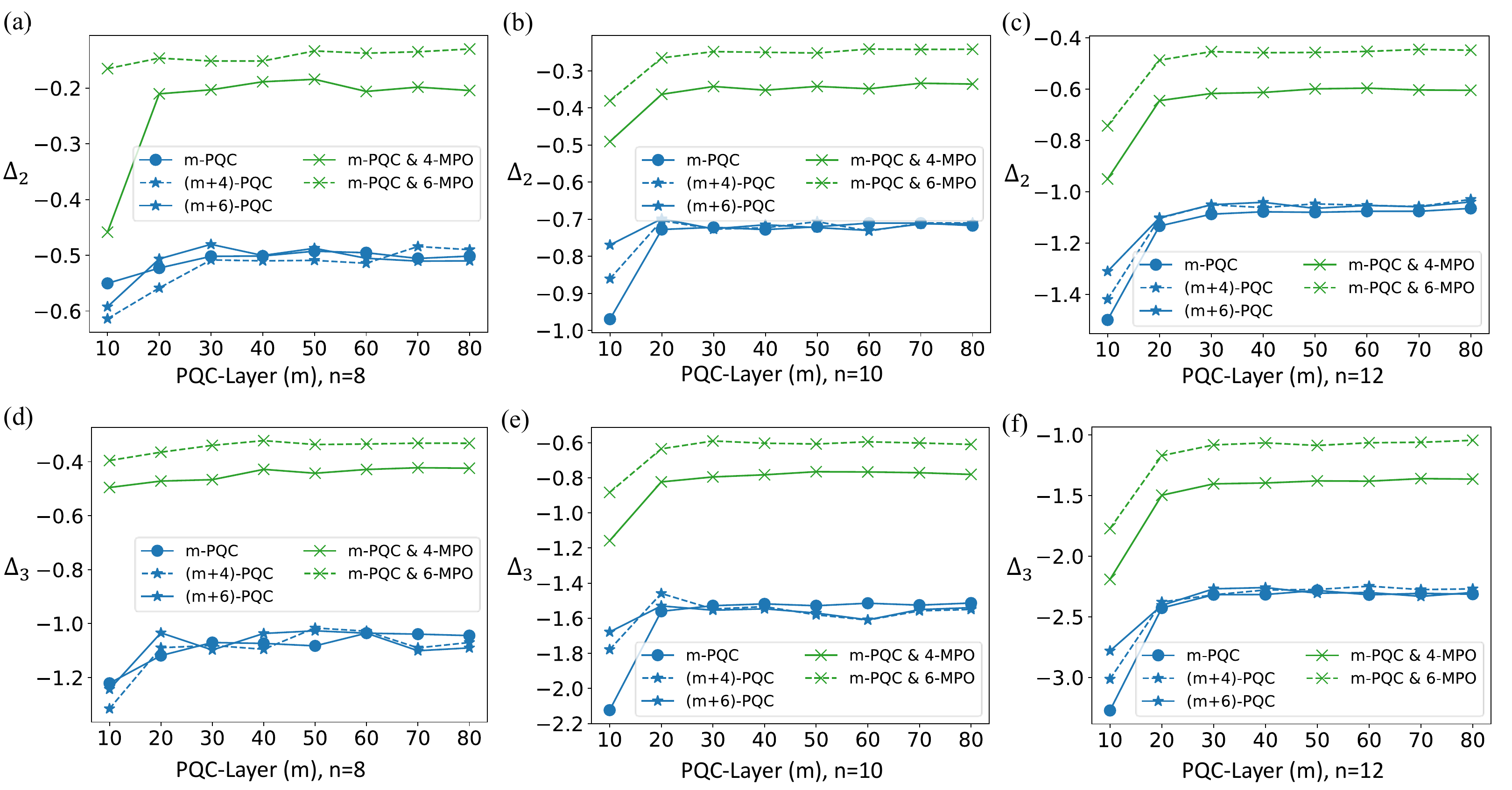}
\caption{{Comparison of the logarithmic difference of the entanglement entropy $\Delta_t$ [defined in Eq.~\eqref{eq:expressivity}] for conventional parametrized quantum circuits and our matrix product operator assisted PQC. 
  Here $\Delta_t$ serves as an indicator of the expressive power of quantum circuits. 
  The following denotations are used: 
  $m$-PQC, a conventional VQE with $m$ layers of PQC;
  ($m+k$)-PQC, a conventional VQE with $m+k$ layers of PQC;
  and $m$-PQC \& $k$-MPO, a TN-PQC structure with $m$ layers of the PQC and $k$ layers of the MPO ($k=4, 6$).
  }
  }
  \label{fig:expressive_power}
\end{figure*}

We first analyze the expressivity of the TN-PQC and VQE methods using the logarithmic difference of entanglement entropy, as defined in Eq.~\eqref{eq:expressivity}.

To conduct our analysis, we choose a layer of parametrized circuit consisting of a layer of Ry$(\theta)$, followed by $n$ layers of controlled-$Z$ gates (CZ$_{i,i+1}$ for $1\leq i\leq n -1$). For the TN-PQC method, we select four or six layers of a parametrized MPO, and the depth of TN-PQC algorithm is represented as the cumulative sum of the PQC and MPO layers.

Figure~\ref{fig:expressive_power} illustrates the logarithmic difference in entanglement entropy, denoted by $\Delta_t$, as the depth of the PQC algorithm increases for both the VQE and TN-PQC algorithms. The numerical results for both the VQE and TN-PQC algorithms, with the depth being $m+4$ (or $m+6$), clearly demonstrate that as the number of PQC layers $m$ increases, the TN-PQC algorithm achieves a significantly larger $\Delta_t$ compared to the VQE algorithm. This observation underscores the enhanced expressiveness of the TN-PQC algorithm and suggests that our approach effectively boosts the expressiveness of shallow quantum circuits through the integration of tensor networks.


On the other hand, it has gradually become recognized that in variational quantum algorithms, high expressivity often brings about the so-called barren plateaus~\cite{holmes2022connecting}. To study this phenomenon, we conducted experiments to observe the gradient of classical and quantum parameters in the TN-PQC algorithm with the depth of the quantum circuit and the size of the quantum system. 
The TN-PQC algorithm cannot completely avoid the barren plateau phenomenon, but the classical part of the hybrid algorithm is more trainable than the corresponding part of the pure quantum algorithm.
A detailed analysis of these observations is provided in Appendix~\ref{app:gradient}.

\subsection{Advantage of the TN-PQC method in accuracy and scalability}
\begin{figure*}[htb]
\centering
  \includegraphics[width=0.75\textwidth]{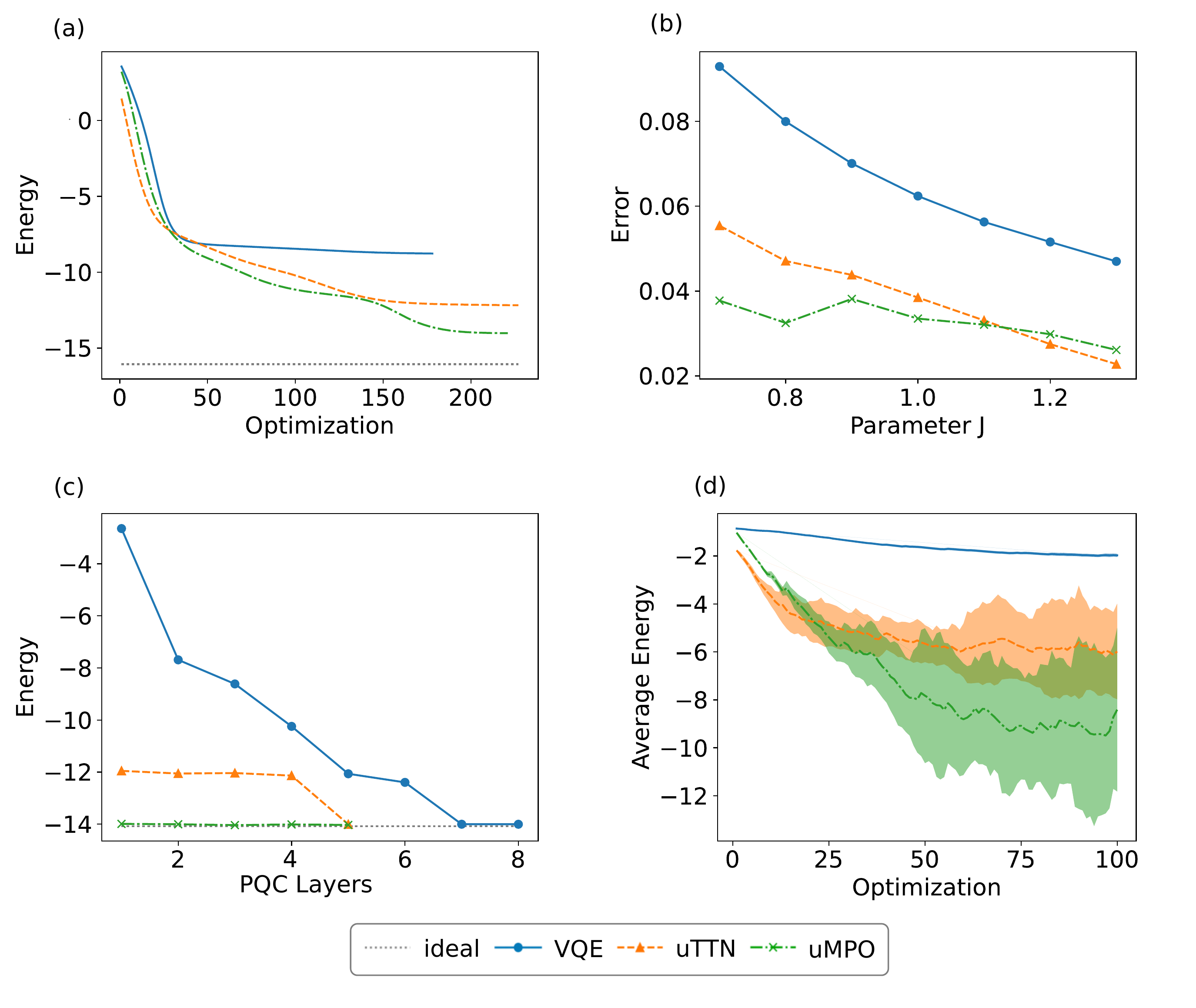}
  \caption{The numerical results demonstrate the superior accuracy and scalability of the TN-PQC structure. The number of layers is chosen to be 2 for uMPO and $\log(n)$ for uTTN with the number of parameters for both being $3(n-1)$. (a) Performance comparison of the pure VQE, uTTN-assisted VQE, and uMPO-assisted VQE on a $(4\times 4)$-qubit 2D transverse field Ising model. All methods have the same PQC, consisting of a layer of parametrized rotations in Pauli $Y$-basis gates and $n$ similar layers of CZ gates for entanglement. (b) Lowest-energy error achieved by the pure VQE, uTTN-assisted VQE, and uMPO-assisted VQE as the time-crystal model parameter $J$ varies. All methods have the same PQC, consisting of 2 layers of parametrized rotation in Pauli $Y$-basis gates and Pauli $X$-basis gates, and $n$ layers of CNOT gates (CNOT$_{i,i+1}$ for $1\leq i\leq n -1$) to induce more entanglement. (c) Estimated ground state energy of a 16-qubit time-crystal model using different algorithms with an increasing number of layers of the PQC. All methods have the same PQC, consisting of a layer of rotation in $Y$-basis gates, followed by $n$ layers of CNOT gates (CNOT$_{i,i+1}$ for $1\leq i\leq n -1$). (d) Estimated energy after $1-100$ optimization steps for the iterative experiment of the pure VQE, uTTN-assisted VQE, and uMPO-assisted VQE on a noisy 16-qubit time-crystal model. The same PQC is maintained in various methods as in (c).}
  \label{fig:numerical}
\end{figure*}

We test the efficacy of the TN-PQC method for determining the ground states of 2D spin-lattice systems with nearest-neighbor interactions, specifically the Ising model. 
The Ising model Hamiltonian is generally given by 
\begin{equation}
    H=-\sum_{\langle i j\rangle} J_{i j} \sigma_i \sigma_j- \sum_j g_j \sigma_j,
\end{equation} 
where $\langle i j\rangle$ denotes the summation over nearest neighbors and $g$ represents the interaction strength between the system and the external magnetic field. 
We choose the Hamiltonian as $H=-J\sum_{\langle i j\rangle} Z_i Z_j-g \sum_j X_j$, which is the well-known transverse field Ising model, where $Z_i$ and $X_j$ are local Pauli operators. 
The parameter set we use is $\left\{J=0.1, g=1\right\}$, and numerical experiments are performed on a $4\times4$ qubit 2D Ising model.
We experimentally compare the performance of the VQE-only circuit, the TN-PQC algorithm with the TN being a uTTN, and the TN-PQC algorithm with the TN being a uMPO (two layers) in determining the ground state of these models, as shown in Fig.~\ref{fig:numerical}(a). All of these algorithms have the same parametrized quantum circuit, consisting of a layer of parametrized rotations in Pauli-$Y$-basis gates and $n$ similar layers of CZ gates for entanglement, as before.

We optimize these circuits using the gradient descent algorithm and set the convergence threshold of energy as $10^{-3}$. 
We find that the TN-PQC algorithm has a much better convergence estimation value of the ground energy, and the TN-PQC algorithm with the TN being a uMPO performs better than a uTTN. 
This suggests that MPOs may be more appropriate for the Hamiltonians with the spin-lattice model.
We compare the performance of the VQE, the TN-PQC algorithm with a uTTN, and the TN-PQC algorithm with an MPO for various parameter models of the time-crystal Hamiltonian, which is given by 
\begin{equation}
    H =-\sum_k\left(J_k Z_{k-1} X_k Z_{k+1}+V_k X_k X_{k+1}+h_k X_k\right).
\end{equation}
In particular, we vary the parameter $J$ from 0.7 to 1.3 with fixed $V = 0.1$ and $h = 0.1$, as shown in Fig.~\ref{fig:numerical}(b). 
To handle this more complex Hamiltonian, we use a deeper circuit consisting of two layers of parametrized rotation in Pauli $Y$-basis gates and Pauli $X$-basis gates, and $n$ layers of controlled-NOT (CNOT) gates (CNOT$_{i,i+1}$ for $1\leq i\leq n -1$) to induce more entanglement.
The figure demonstrates that the TN-PQC method maintains a distinct advantage as $J$ varies. 

While previous experiments have shown the advantages of the TN-PQC method over other methods, these experiments generally used shallow \textit{Ansätze} and did not achieve high accuracy in the ground state energy estimation.
To better understand the capabilities of different approaches in improving accuracy, we investigate the achievable energy accuracy by increasing the number of layers in a parametrized quantum circuit with repeating structures.
We employ the 1D 16-qubit time crystal Hamiltonian to compare the ability of different methods to reduce the number of layers in parametrized quantum circuits.
We set the parameter values to $\left\{J=1, V=0.1, g=0.1\right\}$ and use a layer of rotation in $Y$-basis gates, followed by $n$ layers of CNOT gates (CNOT$_{i,i+1}$ for $1\leq i\leq n -1$) in our parametric quantum circuit.
Figure~\ref{fig:numerical}(c) shows the estimated ground energy for different algorithms as the number of layers increases.
Our results demonstrate that a pure VQE circuit with seven repetitions of the original structure performs as well as the uTTN-assisted VQE circuit with five repetitions and the MPO-assisted original VQE circuit, with all three approaches approaching the theoretical value of the time-crystal Hamiltonian ground-state energy within an error of $1\times10^{-2}$.

\subsection{Robustness of the TN-PQC method under noise}
We assess the robustness of the TN-PQC method to noise by introducing depolarization noise to the single-qubit and two-qubit gates. Multiple fixed 100-step iterative experiments are conducted on the same initial quantum state to obtain the estimation and error bar. The error is estimated using the formula $\varepsilon=3\sqrt{ \sum_{i=1}^S\left (v_i-\bar{v}\right)^2/S^2}$, where $v_i$ denotes the $i$th estimate, $\bar{v}$ denotes the mean, and $S$ denotes the total number of times the experiment is performed.

In practice, the depolarization noise probability for the single-qubit and two-qubit gates (specifically, the rotation Y and CNOT gate) is set at $2\times10^{-5}$ and $5\times10^{-5}$, respectively. The same set of parameters is chosen, and the experiment is repeated $S=40$ times. The mean values and error bars are calculated for each step. Figure~\ref{fig:numerical}(d) shows the estimation results with increasing optimization steps. We observe that while the TN assistance amplifies the noise fluctuation, the TN-PQC algorithm still outperforms the pure VQE even when considering the effect of the estimation error bar.

\section{Discussion}
In this work, we proposed a hybrid framework that combines the strengths of tensor networks and quantum circuits for quantum variational simulation algorithms. 
By incorporating tensor networks with logarithmic or constant layers into the classical optimization process, we demonstrated that the expressiveness of shallow quantum circuits can be significantly enhanced without compromising the accuracy of the simulation. 
Our numerical experiments also showed that the proposed framework is robust to noise, highlighting its potential for practical applications in noisy quantum devices.
We also discussed the general requirements for choosing tensor networks in this class of hybrid frameworks. 
As future work, we envision the possibility of combining tensor networks with other quantum simulation algorithms, such as imaginary time evolution algorithms \cite{mcardle_variational_2019}. 
Moreover, we expect that similar hybrid frameworks can be developed by incorporating classical methods, such as neural networks \cite{carleo_solving_2017,abbas2021power} and quantum Monte Carlo algorithms \cite{foulkes_quantum_2001,carlson2015quantum}, into quantum algorithms. 
We hope that our work can inspire further studies in this direction.
Our work utilizes quantum error correction \cite{endo_hybrid_2021} to push recent quantum hardware to solve static and dynamic quantum problems, as well as quantum optimization \cite{bravyi_classical_2021, cerezo2021variational} for a diverse range of non-trivial applications. 
By leveraging the complementary strengths of tensor networks and quantum circuits, our hybrid framework provides a promising approach for solving challenging quantum problems in a noisy, near-term quantum computing era.


\begin{acknowledgments}
This work was supported by the National Natural Science Foundation of China under Grants No.~12175003 and No.~12147133. 
We acknowledge the High-Performance Computing Platform of Peking University for providing the computational resources required for the numerical simulations conducted in this work.
\end{acknowledgments}

\nocite{*}

\bibliographystyle{apsrev4-2}
\bibliographystyle{unsrt}
\bibliography{apssamp}

\appendix

\section{Expressivity evaluation}
\label{app:expressivity}

\begin{figure}[htb]
\centering
  \includegraphics[width=0.49\textwidth]{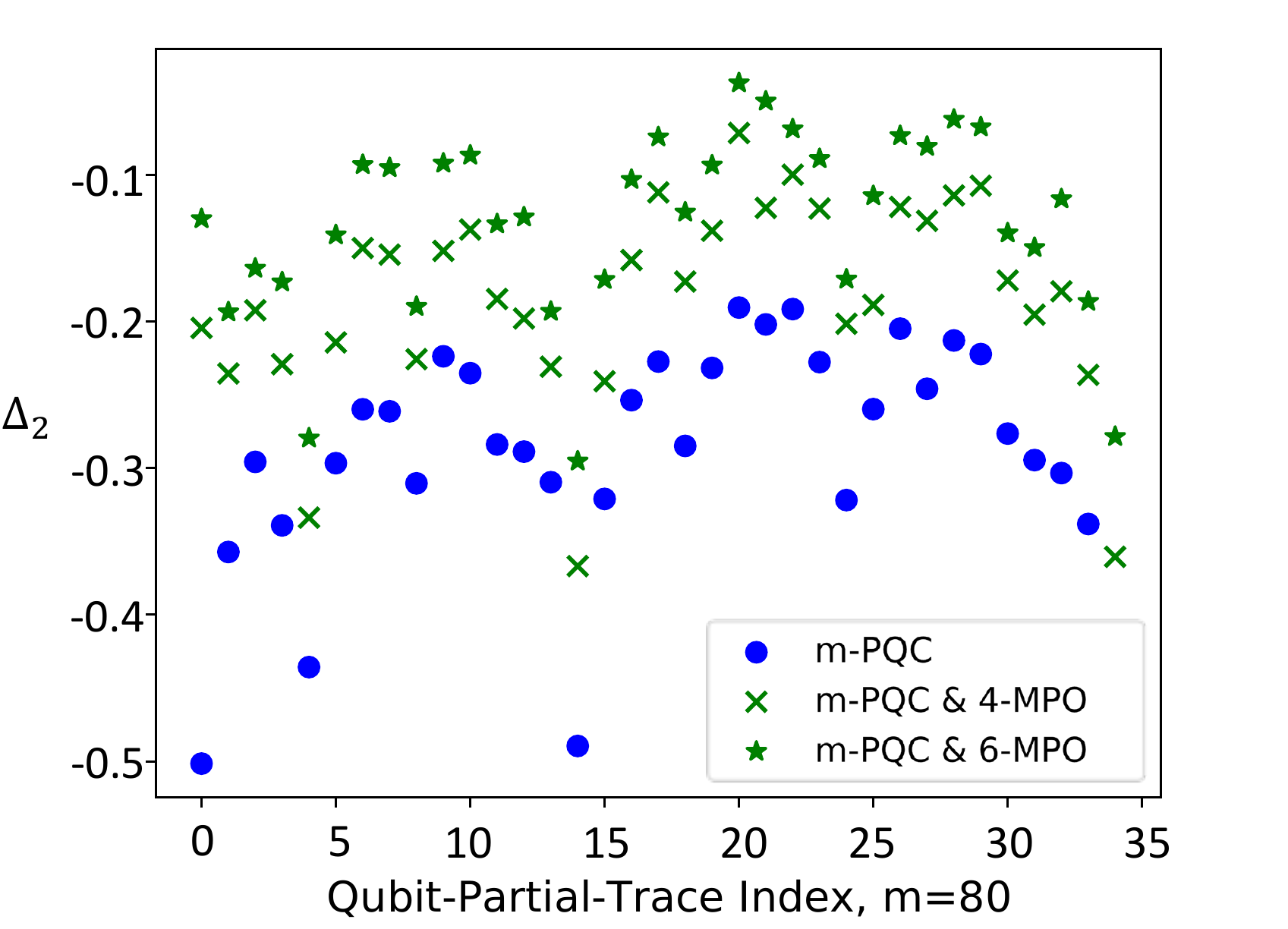}
  \caption{Comparison of the logarithmic differences of entanglement entropy. Here, we randomly discard half of the qubits, and the simulation results indicate that the maximum difference of the metric corresponds to the removal of the first $n/2$ qubits.
  }
  \label{appfig:express}
\end{figure}

The following theorem evaluates if  a distribution is close to the Haar measure by the partial trace. 
\begin{theorem}[\textit{Theorem 15 in Ref. \cite{liu2018entanglement}}]
Let $\Upsilon$ be an $\varepsilon$-approximation $t$-design. Then
\begin{align}
&\Ebb_{\rho\sim \Upsilon}\sbra{\tr\pbra{\rho_A^t}}\leq \Ebb_{\rho \sim \text{Haar}} \sbra{\tr\pbra{\rho_A^t}} + 2^{nt}\varepsilon,\\
&\Ebb_{\rho\sim \Upsilon}\sbra{\log\pbra{\tr\pbra{\rho_A^t}}}\geq \Ebb_{\rho \sim \text{Haar}} \sbra{\log\pbra{\tr\pbra{\rho_A^t} + 2^{nt}\varepsilon}}.
\end{align}
\end{theorem}

We illustrate the logarithmic difference of entanglement entropies for all partitions of a parametrized quantum state generated as in Fig. \ref{fig:expressive_power}(a) in Fig. \ref{appfig:express}.

\section{Tensor-network parametrization}

\begin{figure}[htb]
\centering
  \includegraphics[width=0.5\textwidth]{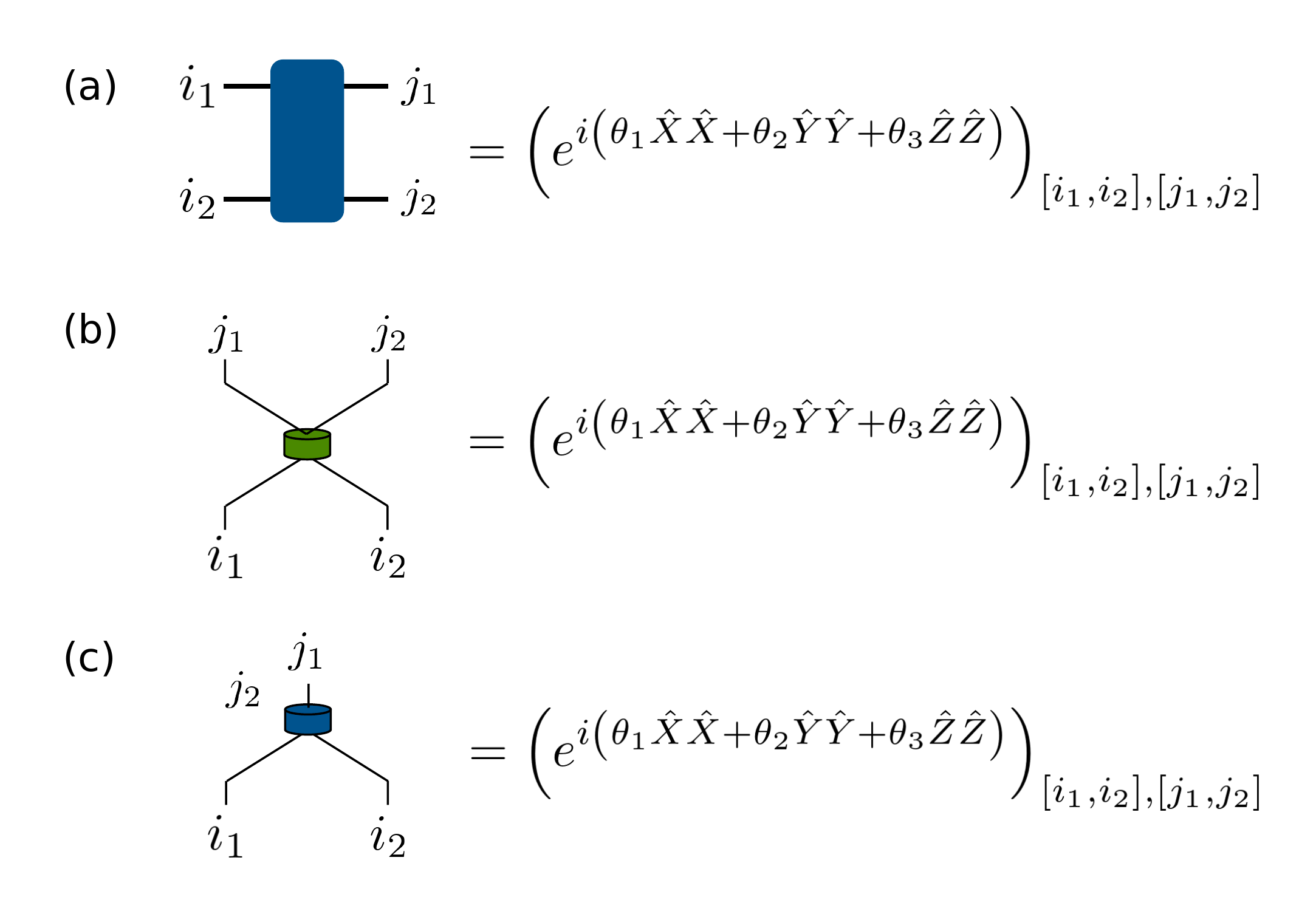}
  \caption{Illustration of how parameters enter the tensor-network building blocks. (a) For the uMPO and uTTN structures as shown in Figs.~\ref{fig:TN}(c) and \ref{fig:TN}(d), there are three tunable parameters in each building block:  $\theta_1,\theta_2,\theta_3$. (b) and (c) For the 2D uMPO and uTTN, as shown in Fig.~\ref{fig:framework}(f), parametrization for each block is similar. In (c) the index $j_2$ is a dangling leg, which we do not draw explicitly. }
  \label{appfig:TN}
\end{figure}

Here, we outline the methodology used to parametrize our unitary tensor networks to ensure the reproducibility of our numerical experiments. Figure~\ref{appfig:TN} illustrates the process of incorporating parameters into the tensor-network building blocks. Specifically, for the uMPO and uTTN structures, as depicted in Figs.~\ref{fig:TN}(c) and \ref{fig:TN}(d), respectively, each building block contains three adjustable parameters denoted by $\theta_1$, $\theta_2$, and $\theta_3$, embedded in the form of
\begin{equation}
    e^{i\left(\theta_1 \hat{X} \hat{X}+\theta_2 \hat{Y} \hat{Y}+\theta_3 \hat{Z} \hat{Z}\right)}.
\end{equation}

Additionally, for the 2D uMPO and uTTN configurations, as shown in Fig.~\ref{fig:framework}(f) of the framework, the parametrization approach for each block is similar. Notably, in Fig.~\ref{fig:framework}(f), the index $j_2$ represents a dangling leg, which is not explicitly drawn for clarity.

By employing this well-defined parametrization scheme, we ensure the consistency and transparency of our tensor-network setup, thereby enabling other researchers to replicate our numerical experiments effectively.

\section{Gradient analysis}
\label{app:gradient}
\begin{figure}[ht]
\centering
  \includegraphics[width=0.46\textwidth]{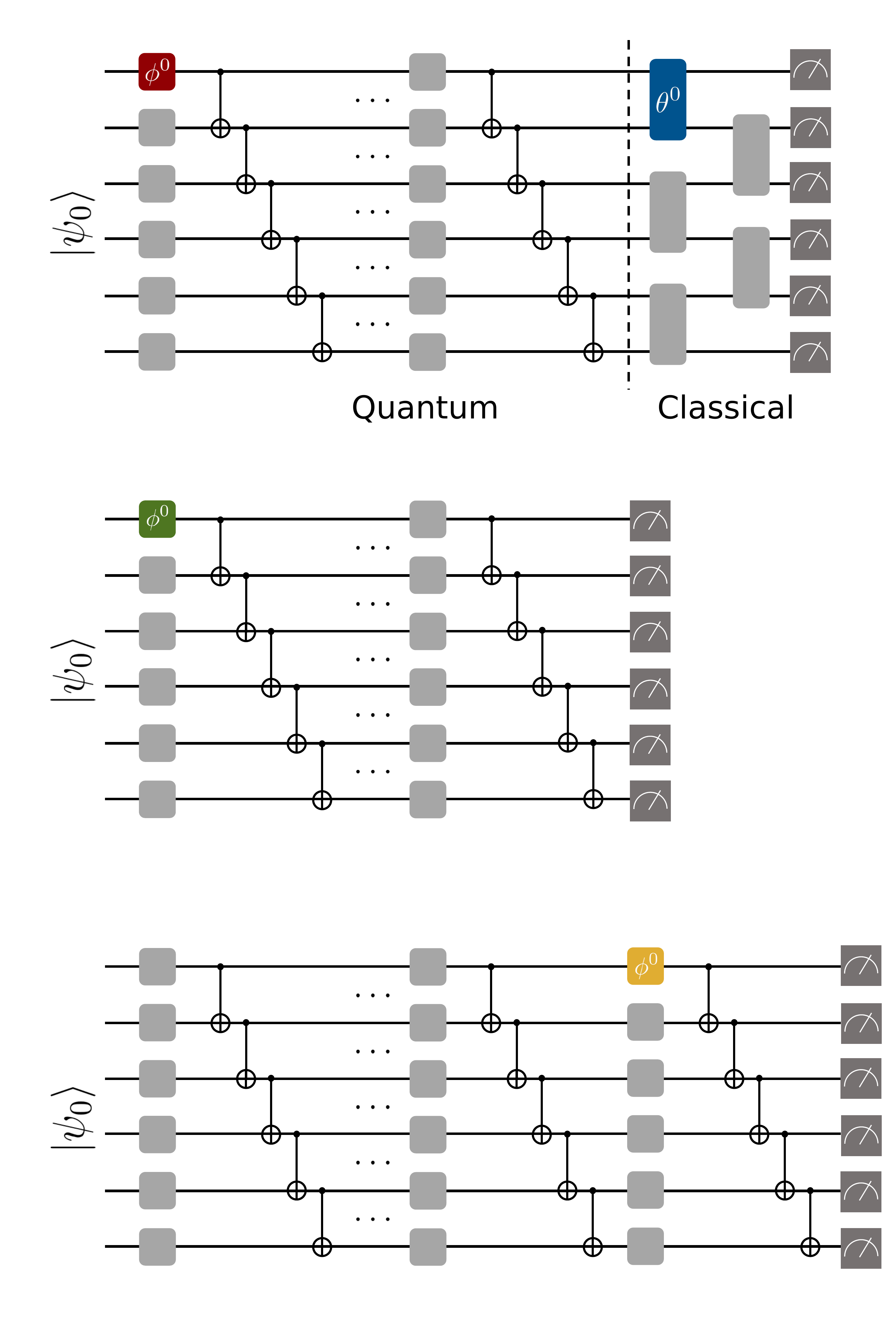}
  \caption{Quantum and classical setups for measuring gradients. Each layer of the quantum circuit consists of single-qubit RX and RY gates and CNOT gates for entanglement, with the classical part being uMPO blocks. Four particular parametrized blocks are identified with different colors indicating the variance analysis of the derivatives of these parameters.}
  \label{appfig:circuit}
\end{figure}
\begin{figure*}[ht]
\centering
  \includegraphics[width=0.8\textwidth]{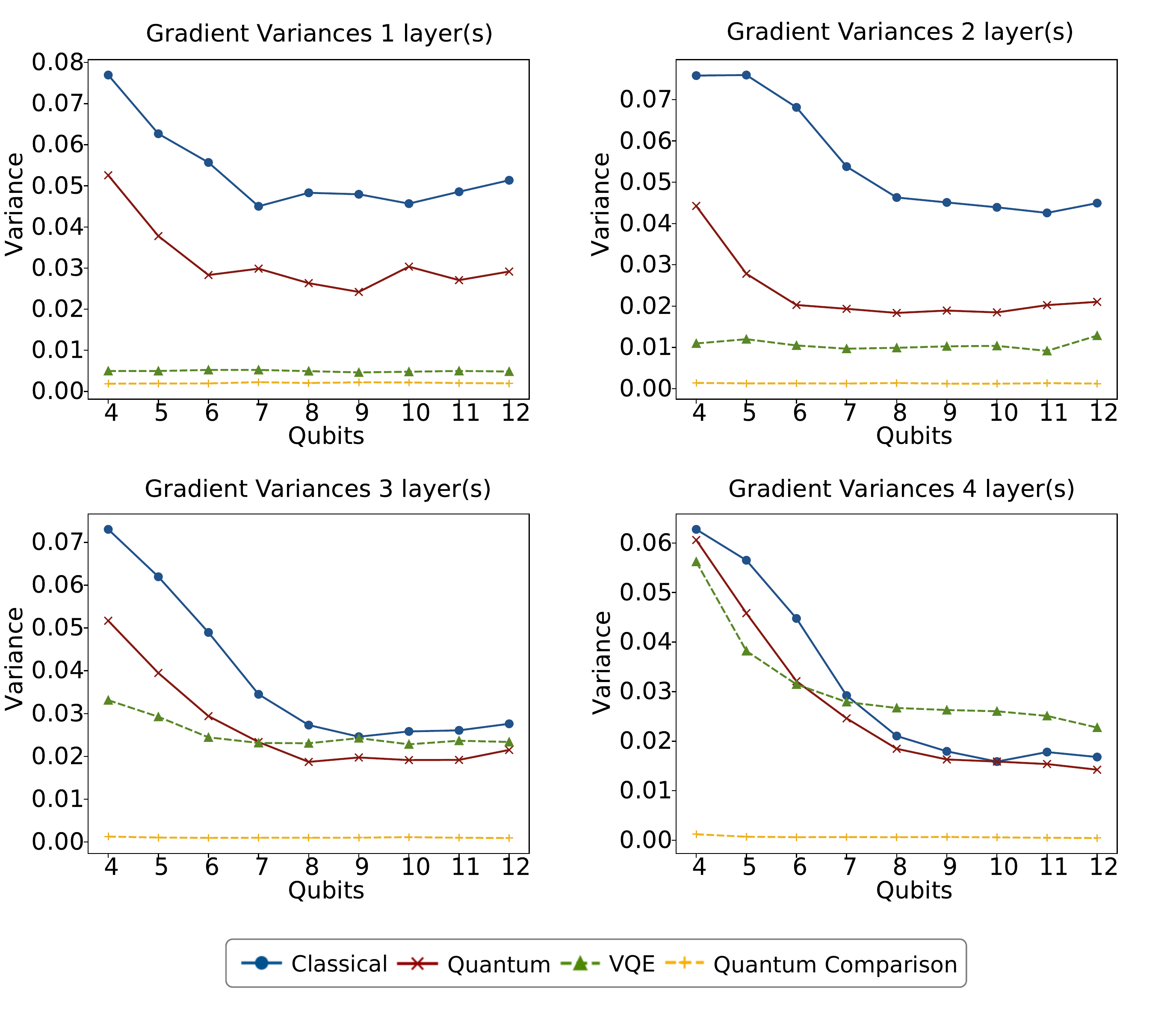}
  \caption{Variances of partial derivatives for different circuit depths and system sizes, where the variance is taken over an ensemble of 1000 unitaries. 
  The colors of the curves correspond to the settings in Fig.~\ref{appfig:circuit}.}
  \label{appfig:gradient}
\end{figure*}

The phenomenon of barren plateaus in variational quantum algorithms has been gradually recognized recently~\cite{holmes2022connecting}. The phenomenon of barren plateaus is recognized by the exponential decay of the gradient of the cost function, denoted by $C$, pertaining to an expressive \textit{Ansatz}, $U(\boldsymbol{\theta})$, as the number of qubits, denoted by $n$, increases.
Here, we have the cost function of the energy expectation form
\begin{equation}
    C(\boldsymbol{\theta})=\operatorname{Tr}\left[H U(\boldsymbol{\theta}) \rho U(\boldsymbol{\theta})^{\dagger}\right],
\end{equation}
where $H$ is the Hamiltonian, $\rho$ is an $n$-qubit input state and $U(\boldsymbol{\theta})$ is a parametrized quantum circuit with parameters $\boldsymbol{\theta}$.

The gradient of the cost function $C$ consists of partial derivatives $\partial_k C:=\partial C(\boldsymbol{\theta}) / \partial \theta_k$, while for a universal \textit{Ansatz} the average of partial derivatives $\partial_k C$ with respect to all parameters $\boldsymbol{\theta}$ should vanish
\begin{equation}
    \left\langle\partial_k C\right\rangle=0 \quad \forall k,
\end{equation}
due to the fact that the \textit{Ansatz} is uniform in any gradient direction.

However, an unbiased unitary \textit{Ansatz} does not necessarily mean that it is untrainable. If one considers the probabilistic definition, i.e., the gradient has a certain probability of deviating from the mean value $0$, this comes from the Chebyshev inequality
\begin{equation}
    \Pr\left(\left|\partial_k C\right| \geqslant \delta\right)\leqslant \frac{\operatorname{Var}\left(\partial_k C\right)}{\delta^2},
\end{equation}
where we define the variance as
\begin{equation}
    \operatorname{Var}\left(\partial_k C\right)=\left\langle\left(\partial_k C\right)^2\right\rangle-\left\langle\partial_k C\right\rangle^2,
\end{equation}
in which the average is over all possible parameters $\boldsymbol{\theta}$.

In order to observe the phenomenon of barren plateaus in our TN-PQC \textit{Ansatz}, we experimentally obtained the derivatives of the first quantum and classical parameters of their first layer, as shown in Fig.~\ref{appfig:circuit}, and calculated their variance from 1000 randomly and uniformly chosen parameters, varying with the number of quantum line layers and the size of the system.  
As a comparison, we further investigate whether (i) the inclusion of a tensor-network would alleviate the problem of vanishing derivatives of the quantum parameters and (ii) the derivatives of a parametrized classical tensor-network would be more trainable than the same quantum parameters. Hence we also collected the variance of the derivatives of the quantum parameters for the VQE circuit with the tensor-network portion removed and the variance of the derivatives of the parameters of the quantum circuit with the classical portion replaced by the quantum part with the same number of parameters, as illustrated in Fig.~\ref{appfig:circuit}.

As shown in Fig.~\ref{appfig:gradient},  we find that the variance curve for quantum circuits with shallow layers has a flat long tail as the number of qubits increases, and the curve gradually adheres to an exponentially vanishing curve as the number of quantum circuit layers increases. 
This suggests that although the TN-PQC algorithm can improve the expressivity of \textit{Ansätze} to a certain extent, it is still not able to completely avoid the barren plateau phenomenon as a variational quantum algorithm. 

Nonetheless, the combination of the PQC and TN alleviates the awkwardness of disappearing gradients in some ways. In the shallow layers, when combining the quantum circuit with tensor-networks, there is a certain advantage in terms of the variance of quantum parameter gradients compared to the original quantum setting. However, as the quantum depth increases, this advantage gradually diminishes. Furthermore, the variance of the classical part of the parameters with overwhelmingly large derivatives with respect to the quantum parameters of the same parameters at the same positions seems to imply a potential advantage of combining tensor-networks with parametrized quantum circuits.

\end{document}